\documentclass[
    ,final            
  ]
  {aipproc}

\layoutstyle{6x9}

\begin{document}

\title{HERMES high-resolution spectroscopy of HD 149382 \\
Where did the planet go?}

\classification{97.82.Fs}
\keywords      {Substellar companions; planets }

\author{V.A. Jacobs}{
  address={Instituut voor Sterrenkunde, Katholieke Universiteit Leuven, Celestijnenlaan 200D, B-3001 Leuven, Belgium}
}

\author{R.H. \O stensen}{
  address={Instituut voor Sterrenkunde, Katholieke Universiteit Leuven, Celestijnenlaan 200D, B-3001 Leuven, Belgium}
}

\author{H. Van Winckel}{
  address={Instituut voor Sterrenkunde, Katholieke Universiteit Leuven, Celestijnenlaan 200D, B-3001 Leuven, Belgium}
}

\author{S. Bloemen}{
  address={Instituut voor Sterrenkunde, Katholieke Universiteit Leuven, Celestijnenlaan 200D, B-3001 Leuven, Belgium}
}

\author{P.I. P\'{a}pics}{
  address={Instituut voor Sterrenkunde, Katholieke Universiteit Leuven, Celestijnenlaan 200D, B-3001 Leuven, Belgium}
}

\author{G. Raskin}{
  address={Instituut voor Sterrenkunde, Katholieke Universiteit Leuven, Celestijnenlaan 200D, B-3001 Leuven, Belgium}
}

\author{J. Debosscher}{
  address={Instituut voor Sterrenkunde, Katholieke Universiteit Leuven, Celestijnenlaan 200D, B-3001 Leuven, Belgium}
}

\author{S. Uttenthaler}{
  address={Instituut voor Sterrenkunde, Katholieke Universiteit Leuven, Celestijnenlaan 200D, B-3001 Leuven, Belgium}
}
\author{E. Van Aarle}{
  address={Instituut voor Sterrenkunde, Katholieke Universiteit Leuven, Celestijnenlaan 200D, B-3001 Leuven, Belgium}
}

\author{C. Waelkens}{
  address={Instituut voor Sterrenkunde, Katholieke Universiteit Leuven, Celestijnenlaan 200D, B-3001 Leuven, Belgium}
}

\author{E. Bauwens}{
  address={Instituut voor Sterrenkunde, Katholieke Universiteit Leuven, Celestijnenlaan 200D, B-3001 Leuven, Belgium}
}

\author{T. Verhoelst}{
  address={Instituut voor Sterrenkunde, Katholieke Universiteit Leuven, Celestijnenlaan 200D, B-3001 Leuven, Belgium}
}

\author{C. Gielen}{
  address={Instituut voor Sterrenkunde, Katholieke Universiteit Leuven, Celestijnenlaan 200D, B-3001 Leuven, Belgium}
}

\author{H. Lehmann}{
  address={Th\"uringer Landessternwarte Tautenburg, Karl-Schwarzschild-Observatorium, 07778 Tautenburg, Germany}
}

\author{R. Oreiro}{
  address={Instituut voor Sterrenkunde, Katholieke Universiteit Leuven, Celestijnenlaan 200D, B-3001 Leuven, Belgium}
  ,altaddress={Instituto de Astrof\'{i}sica de Andaluc\'{i}a-CSIC, E-18008 Granada, Spain} 
}

\begin{abstract}
A close substellar companion has been claimed to orbit the bright sdB star HD $149382$ with
a period of $2.391$ d. In order to check this important discovery we gathered $26$ high resolution
spectra over $55$ days with the {\sc hermes} spectrograph on the $1.2$m Mercator
telescope on La Palma, and analyzed the resulting radial velocities. Our data show no sign
of any significant radial-velocity periodicities, and from the high precision of our measurements
we rule out any RV variations with amplitudes higher than $0.79$ km/s on periods
shorter than $50$ days.
\end{abstract}

\maketitle


\section{Introduction}
\begin{figure}
  \includegraphics[height=.3\textheight]{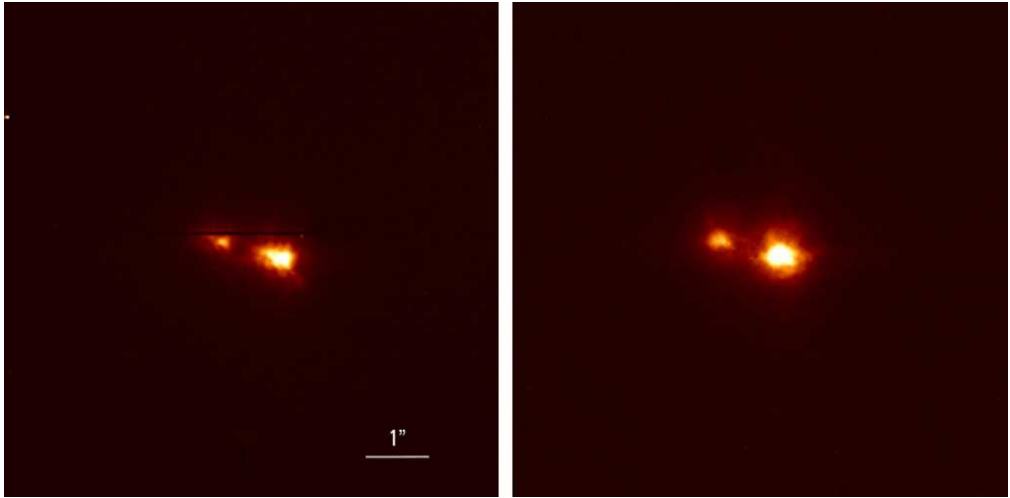}
  \caption{A red companion was clearly detected about 1" away from the
subdwarf in high resolution H-band imaging obtained with the {\sc naomi} adaptive
optics system at the William Herschel Telescope on La Palma, by
\cite{roy2005}}
 \label{Ingrid}
\end{figure}

HD 149382 was recognised to be a subdwarf B star by \cite{MacConnell1972},
and the spectroscopic analysis of \cite{saffer1994} place it
on the hot end of the extreme horizontal branch (EHB). With a magnitude
of V\,=\,8.9 it is the brightest EHB star in the sky, and one of just a
handful of such subdwarfs that can be easily observed with
high-resolution spectroscopy on 1-m class telescopes.
The first hints of a binary nature for HD\,149382 came with the study of
\cite{ulla1998}, where they used infrared $JHK$ colours as an
indicator of binarity, and concluded that HD\,149382 has a very
signficant IR excess, compatible with a companion of class K1 to G2.
Such binaries are predicted to form via stable Roche lobe overflow, and
result in orbital periods of hundreds of days \citep{Han2003}. A red
companion was clearly detected about 1" away from the subdwarf in high
resolution H-band imaging obtained with the {\sc naomi} adaptive optics system
at the William Herschel Telescope on La Palma, by \cite{roy2005}
(see Fig. \ref{Ingrid}).
At $\sim$75\,pc, this separation corresponds to $\sim$75\,AU making
it too wide for the two stars to have interacted during the evolution of
the primary. There is also no significant change in the separation
between the images obtained in 2002 and 2004, so any orbital period must
be on the order of decades, if these objects are gravitationally bound
at all.

Later, a close companion was claimed to be
found around HD\,149382 by \cite{geier2009}, based on 15 high-resolution
spectra ($R$\,=\,30000--48000) on three different spectrographs
(ESO-2.2m/{\sc feros}, CAHA-2.2m/{\sc foces}, and McDonald-2.7m/Coud\'e)
taken within four years and one
additional VLT/{\sc uves} spectrum ($R$\,=\,80000). This was also the first claim that a
close substellar companion, able to influence the stellar evolution of the host
star, was found, thus challenging evolutionary experts to come up
with a theory in which single stars with substellar companions were
able to produce a subdwarf. This led to the profound question of
what the actual influence of the companion might be, as such a low
mass companion would be able to aid the formation of a hot subdwarf.

\section{Observations}
\begin{table}
\begin{tabular}{lrrrr}
\hline
  \tablehead{1}{l}{b}{Date}
  & \tablehead{1}{r}{b}{Telescope/Instrument}
  & \tablehead{1}{r}{b}{$\rm N_{exp}$}
  & \tablehead{1}{r}{b}{$\rm T_{exp}$ $[s]$}
  & \tablehead{1}{r}{b}{Observer}   \\
\hline
$2002/08/16    $& WHT/{\sc naomi}/{\sc ingrid}  &  $14 $      & $  5.0 $ &R.H.\O     \\
$2004/06/29    $& WHT/{\sc naomi}/{\sc ingrid}  &  $ 2 $      & $  3.0 $ &R.H.\O     \\ \hline
$2009/06/30    $& Mercator/{\sc hermes}   &  $ 1 $      & $  1500 $ &R.O.     \\
$2009/06/30    $& Mercator/{\sc hermes}   &  $ 1 $      & $  547 $ &J.D.     \\
$2009/07/12    $& Mercator/{\sc hermes}   &  $ 1 $      & $  900 $ &S.U.     \\
$2009/07/13--18$& Mercator/{\sc hermes}   &  $ 9 $      & $ 1800 $ &E.v.A.     \\
$2009/07/21    $& Mercator/{\sc hermes}   &  $ 1 $      & $ 1800 $ &C.W.     \\
$2009/07/31    $& Mercator/{\sc hermes}   &  $ 1 $      & $  900 $ &E.B.     \\
$2009/08/06--13$& Mercator/{\sc hermes}   &  $ 8 $      & $ 1\times615,\, 1\times900,\, 6\times1800 $ &T.V.     \\
$2009/08/14--15$& Mercator/{\sc hermes}   &  $ 2 $      & $ 1800 $ &C.G.     \\\hline
\end{tabular}
\caption{Overview of the observations of HD\,149382. The first two rows
refer to the high resolution infrared imaging with the {\sc naomi}
adaptive optics system at the William Herschel Telescope on La Palma,
while the other rows refer to the high resolution {\sc hermes} spectra,
taken with the Mercator telescope at La Palma. For completeness,
the number of exposures, the exposure times and the observers are listed.}
\label{Table_Observers}
\end{table}

We gathered 26 high-resolution spectra over a timebase of 55 days
(Table \ref{Table_HERMES_HD}) with the {\sc hermes} spectrograph on the 1.2-m
Mercator telescope on La Palma. {\sc hermes} reaches a
spectral resolution of $\sim$85\,000 and a spectral coverage from 377 to 900
nm in a single exposure with a peak efficiency of 28\%. The spectrograph
is bench-mounted and fibre-fed, and resides in a temperature and
pressure controlled enclosure to guarantee instrumental stability. The
high resolution fibre has a sky aperture of 2.5 arcsec. The red
companion to HD\,149382 (see Fig. \ref{Ingrid}) is therefore close enough to
enter the aperture and slightly contaminate the spectrum of the target.

Some spectra were taken during the early testing of the spectrograph.
For three spectra, a problem occurred during integration and the
exposure was read out early. The resulting spectra were of too low
signal-to-noise to be useful, leaving us with 23 spectra to analyse.
Four HeI lines at 6678.15\,\AA, 5015.68\,\AA, 4713.14\,\AA\, and 5875.6251\,
\AA, were used to determine the radial velocities by simultaneous fitting
using the software package Molly\footnote{We thank Tom Marsh for the use
of 'molly'.}. One spectrum resulted in an error of
more than 2 km/s and as this was much higher than the other errors (at a
3$\sigma$-level), it was deleted from the list. 22 useful spectra remain
in Table. \ref{Table_HERMES_HD}. 
\begin{figure}
  \includegraphics[width=10cm,angle=-90,clip]{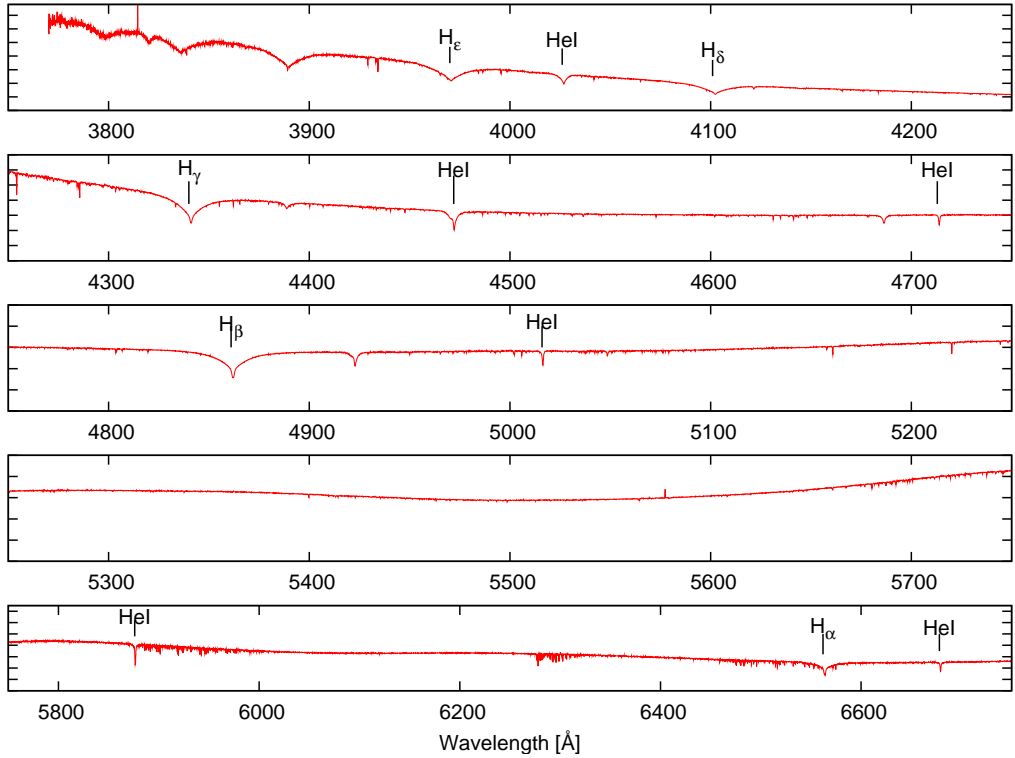}
  \caption{Median spectrum of 22 observations of HD 149382. The four helium lines, that were used for the determination
of the radial velocities are indicated.}
 \label{Spectrum}
\end{figure}

\begin{table}
\begin{tabular}{cccc}
\hline
  \tablehead{1}{r}{b}{Date}
  & \tablehead{1}{r}{b}{RV [km/s]}
  & \tablehead{1}{r}{b}{Date}
  & \tablehead{1}{r}{b}{RV [km/s]} \\ \hline
2455003.45024 & 26.68$\pm$0.15 & 2455034.45301 & 26.28$\pm$0.13 \\
2455025.49846 & 26.09$\pm$0.19 & 2455050.41444 & 25.13$\pm$0.22 \\
2455026.40797 & 26.05$\pm$0.11 & 2455051.38909 & 26.15$\pm$0.11 \\
2455027.41298 & 26.42$\pm$0.10 & 2455051.51929 & 26.09$\pm$0.11 \\
2455027.55091 & 26.48$\pm$0.11 & 2455055.41469 & 25.75$\pm$0.19 \\
2455028.43085 & 26.36$\pm$0.10 & 2455055.53976 & 26.37$\pm$0.10 \\
2455028.59094 & 25.96$\pm$0.13 & 2455056.38705 & 26.33$\pm$0.14 \\
2455029.40622 & 26.21$\pm$0.11 & 2455056.43345 & 26.48$\pm$0.13 \\
2455030.41048 & 26.16$\pm$0.11 & 2455057.41243 & 26.61$\pm$0.17 \\
2455030.50488 & 25.93$\pm$0.12 & 2455058.38674 & 26.70$\pm$0.11 \\
2455031.41011 & 26.38$\pm$0.11 & 2455058.51017 & 26.44$\pm$0.14 \\
\hline
\end{tabular}
\caption{Overview of the 22 radial velocities of HD 149382 as derived from the {\sc hermes} spectra.}
\label{Table_HERMES_HD}
\end{table}

\section{Results}
\begin{figure}
  \includegraphics[width=7cm,clip]{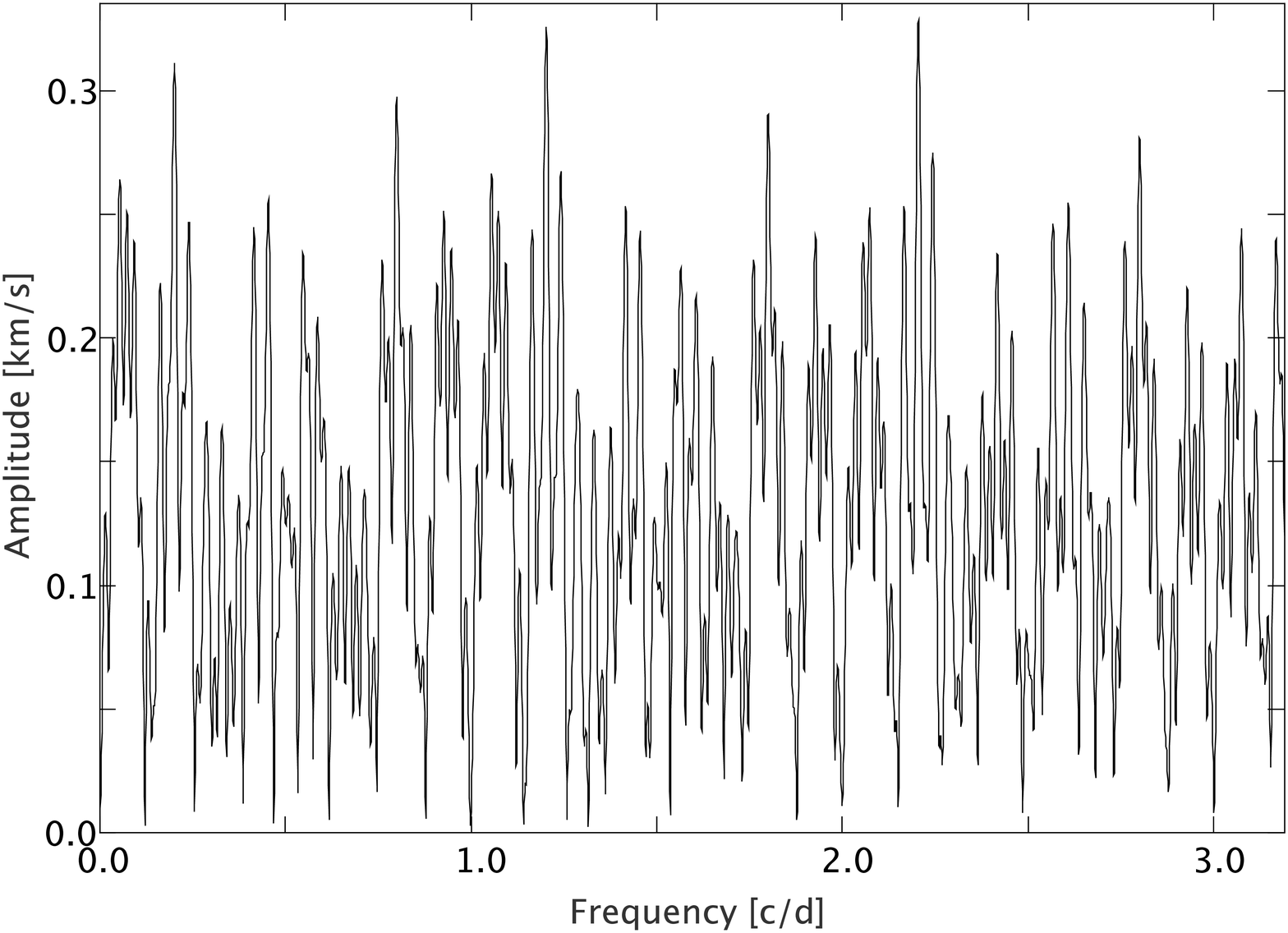}
  \includegraphics[width=7cm,clip]{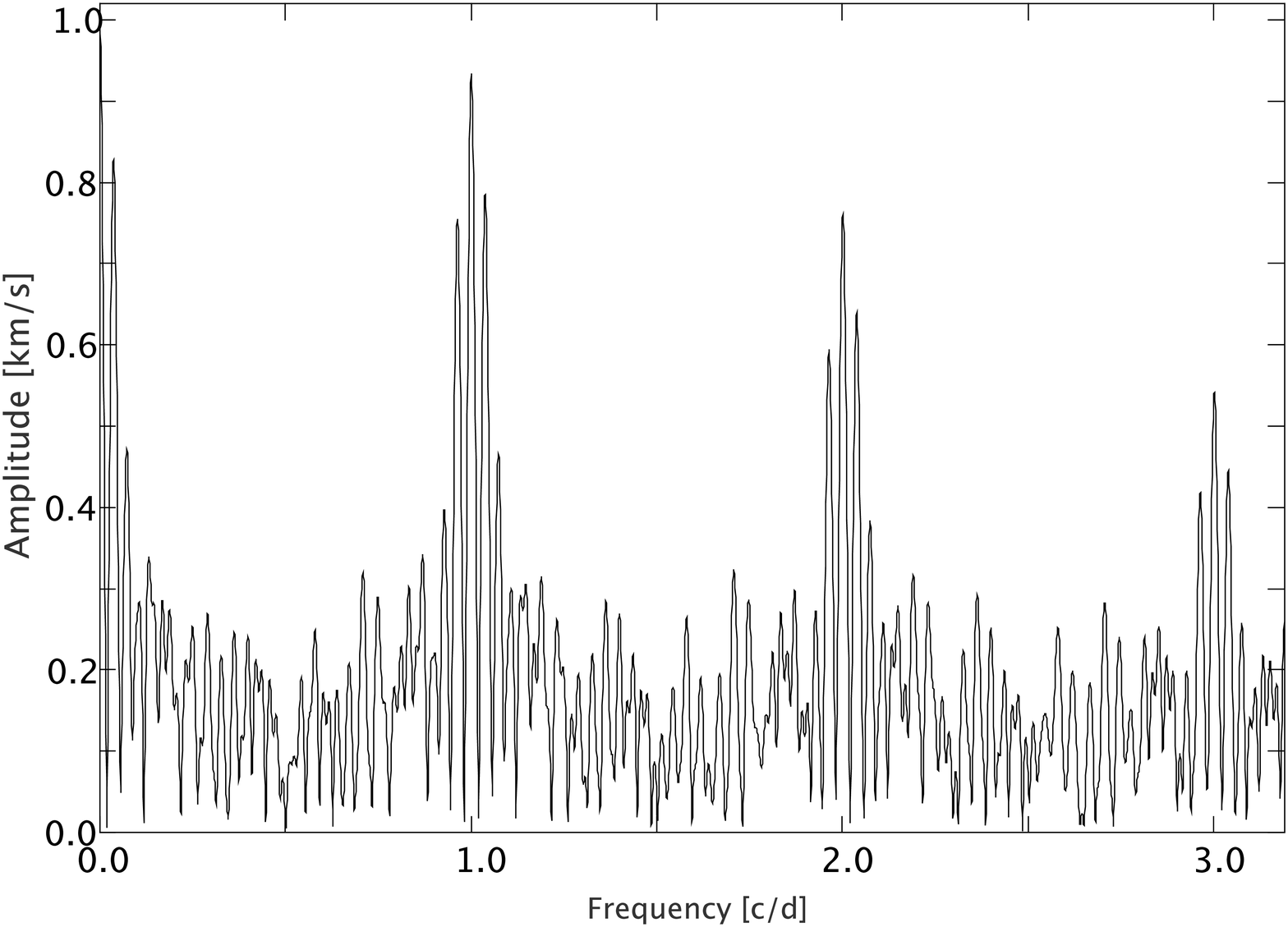}
  \caption{Left panel: The amplitude spectrum of our {\sc hermes} observations.
   Clearly, no significant frequencies can be found, nor do we find similar
   amplitudes to those reported by \cite{geier2009}.
   Right panel: The spectral window of our {\sc hermes} observations clearly
   shows the 1 day alias.}
 \label{AmplitudeSpectrum}
\end{figure}

Fig.~\ref{AmplitudeSpectrum} shows the amplitude spectrum of our {\sc hermes}
observations, and the associated spectral window. Clearly
no significant frequencies can be found, nor do we find similar
amplitudes to those reported by \cite{geier2009}.

\begin{figure}
  \includegraphics[width=10cm]{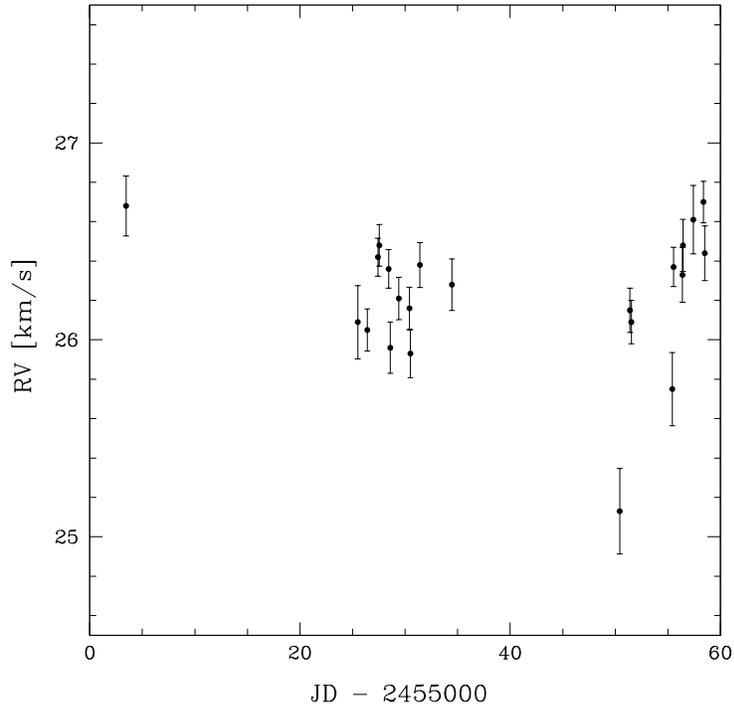}
  \caption{Radial velocities of HD 149382 as derived from the {\sc hermes} spectra.}
 \label{Molly}
\end{figure}

\section{Conclusions}
Our additional data clearly shows that there is as yet no evidence for a
close substellar companion surrounding the bright sdB star HD\,149382.
We gathered 22 spectra with the {\sc hermes} spectrograph on the Mercator Telescope.
Our data show no sign of any radial-velocity periodicities, and
from the high precision of our measurements we rule out any RV
variations with amplitudes
higher than 0.79 km/s on periods shorter than 50 days. 

\section{ONLINE DATA}
The extracted spectra used in this work are available on-line at
\url{http://www.ster.kuleuven.be/instruments/hermes/data/HD149382/}.

\bibliographystyle{aipproc}   
\bibliography{Bib}

\end{document}